\def\cm{{\rm\thinspace cm}}

\def\keV{{\rm\thinspace keV}}
\def\km{{\rm\thinspace km}}
\def\kpc{{\rm\thinspace kpc}}

\def\Mpc{{\rm\thinspace Mpc}}

\def\s{{\rm\thinspace s}}
\def\yr{{\rm\thinspace yr}}

\def\cmsq{\hbox{$\cm^2\,$}}

\def\pcmcu{\hbox{$\cm^{-3}\,$}}

\def\kmps{\hbox{$\km\s^{-1}\,$}}

\def\ps{\hbox{$\s^{-1}\,$}}

\def\kmpspMpc{\hbox{$\kmps\Mpc^{-1}$}}

\documentclass[usegraphicx]{mn2e}

\usepackage{amssymb}
\usepackage{mathptmx}

\include{defn}
\begin{document}

\title[Deep Chandra observation of the Perseus cluster]
{A deep \emph{Chandra} observation of the Perseus cluster: shocks and
ripples}
\author[A.C. Fabian et al]{A.C. Fabian$^1$, J.S. Sanders$^1$, S.W.
Allen$^1$, 
C.S. Crawford$^1$, K. Iwasawa$^1$, \newauthor R.M. Johnstone$^1$, 
R.W. Schmidt$^{1,2}$ and G.B. Taylor$^3$\\
$^1$  Institute of  Astronomy, Madingley Road, Cambridge CB3 0HA\\
$^2$ Institut f\"ur Physik, Universit\"at Potsdam, Am Neuen
Palais 10, 14469 Potsdam, Germany \\
$^3$ National Radio Astronomy Observatory, Socorro, NM 87801, USA }

\maketitle

\begin{abstract}
We present preliminary results from a deep observation lasting almost
200~ks, of the centre of the Perseus cluster of galaxies around
NGC\,1275. The X-ray surface brightness of the intracluster gas beyond
the inner 20~kpc, which contains the inner radio bubbles, is very
smooth apart from some low amplitude quasi-periodic ripples. A clear
density jump at a radius of 24~kpc to the NE, about 10~kpc out from
the bubble rim, appears to be due to a weak shock driven by the
northern radio bubble.  A similar front may exist round both inner
bubbles but is masked elsewhere by rim emission from bright cooler
gas. The continuous blowing of bubbles by the central radio source,
leading to the propagation of weak shocks and viscously-dissipating
sound waves seen as the observed fronts
and ripples, gives a rate of working which balances the radiative
cooling within the inner 50~kpc of the cluster core.
\end{abstract}

\section{Introduction}
The Perseus cluster, A426, is the X-ray brightest cluster of galaxies
in the sky. The giant central galaxy, NGC\,1275, hosts the radio
source 3C84 (Pedlar et al 1990), from which jets have blown two
diametrically-oppposed bubbles in the hot intracluster medium
(B\"ohringer et al 1993; Fabian et al 2000). Surrounding NGC\,1275 is
a spectacular H$\alpha$ nebulosity with filaments extending over
100~kpc (Lynds 1970; Conselice et al 2001).

The radiative cooling time of the hot gas drops inward to
$\sim10^8$~yr around NGC\,1275, and the temperature drops down from
from $\sim 7$ to about 3 keV. As has been found for many clusters with
Chandra and XMM-Newton data (Peterson et al 2001, 2002; Tamura et al
2001), there is little gas at lower temperatures. A standard cooling
flow (Fabian 1994) is not taking place despite the short cooling
times, presumably due to some balancing heat source. Plausible heat
sources are an active galactic nucleus (Tucker \& Rosner 1983, Tabor
\& Binney 1993; Churazov et al 2000; B\"ohringer et al 2002) and the
hot atmosphere surrrounding the AGN (Bertschinger \& Meiksin 1986;
Narayan \& Medvedev 2001). The jets and bubbles may heat the
surrounding gas from the centre (Br\"uggen \& Kaiser 2001; Quilis et
al 2001; Reynolds et al 2001; Basson \& Alexander 2003) or thermal
conduction may heat it from the outside (Voigt et al 2002; Fabian,
Voigt \& Morris 2002a; Zakamska \& Narayan 2003). Combinations of
heating and conduction have also been proposed (Ruszkowski \& Begelman
2002). It has proven difficult in many models to match the observed
temperature profiles (Brighenti \& Mathews 2003). Other possibilities
involving inhomogeneous metallicity (Fabian et al 2001; Morris \&
Fabian 2002) and mixing with cooler gas (Fabian et al 2001; 2002c)
remain.

We observed the core of the Perseus cluster with Chandra for about
25~ks in 2001 (Fabian et al 2000; Schmidt, Fabian \& Sanders 2002).
Here we report on a recent Chandra observation which was almost ten
times deeper ($\sim 200$~ks). The complex two-dimensional temperature
distribution of the hot gas is now clear, and subtle structures are
revealed. These may be dissipating sound waves caused by the bubbles.
The power from the sound waves is consistent with the radiative losses
of the cluster core.

The Perseus cluster is at a redshift of 0.0183. We assume that
$H_0=70\kmpspMpc$ so that 1 kpc corresponds to about 2.7 arcsec.

\section{Analysis}
Two observations of the cluster were made using \emph{Chandra} with
its ACIS-S detector, on 2002 Aug 8 (exposure 95.8~ks) and 10 (exposure
95.4~ks). The datasets show no evidence for contamination by flares
and were merged into a single events file since the nominal
roll angle of the spacecraft was essentially identical. The analysis
presented here was made on this merged events file, which was
processed with the acisD1999-09-16gainN0004 gain file.

The spectral analysis was made on the data from the S3 chip,
although the neighbouring chips were used for imaging.  The datasets
were processed using \textsc{ciao} 2.3, and \textsc{xspec} 11.2 was
used to analyse the spectra.

\begin{figure}
  \centering
  \includegraphics[width=0.99\columnwidth]{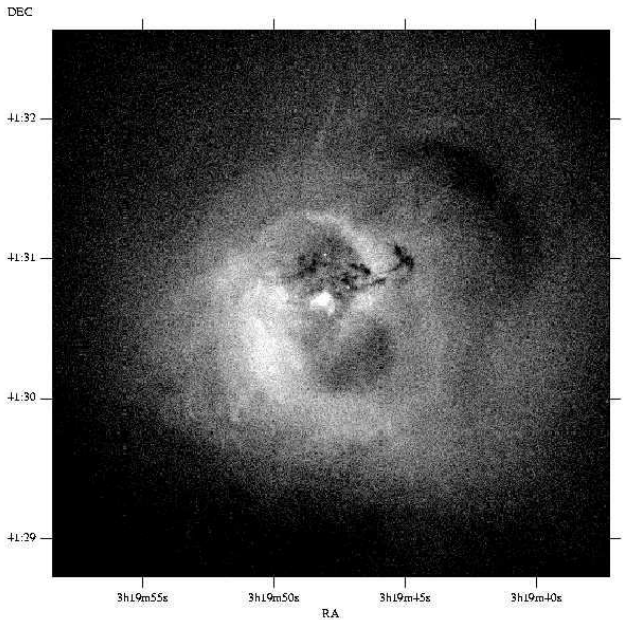}
    \caption{Central regions of the cluster in the 0.3 to 1.5~keV
    band. Pixels are 0.49 arcsec in dimension. North is to the top of
    this image. Absorption from the $8000\kmps$ system projected to
    the N of the nucleus of NGC1275 is evident as are wisps to the N
    and NW which correlate with optical H$\alpha$ filaments.}
    \label{fig:central_lowenergy}
\end{figure}

We show in Fig.~\ref{fig:central_lowenergy} an image of the
0.3-1.5~keV emission from the central $\sim 5.5$~arcmin with 0.49
arcsec pixels. The image was corrected for two linear nodal structures
(at position angle 80 deg near the top and the lower third of the
image), by dividing by an exposure map generated using the
\textsc{ciao} \textsc{merge\_all} script.  Similar images in the
1.5--3.5 and 3.5--7~keV bands are shown in Fig.~2, binned on 0.98 and
1.96 arcsec pixels respectively. The radio bubbles are surrounded by a
$\sim 10\kpc$ thick region of hard X-ray emission (Fig.~3b) and by
structured softer patches and rims (Figs.~1 and 8). 


\begin{figure}
  \centering
  \includegraphics[width=0.99\columnwidth]{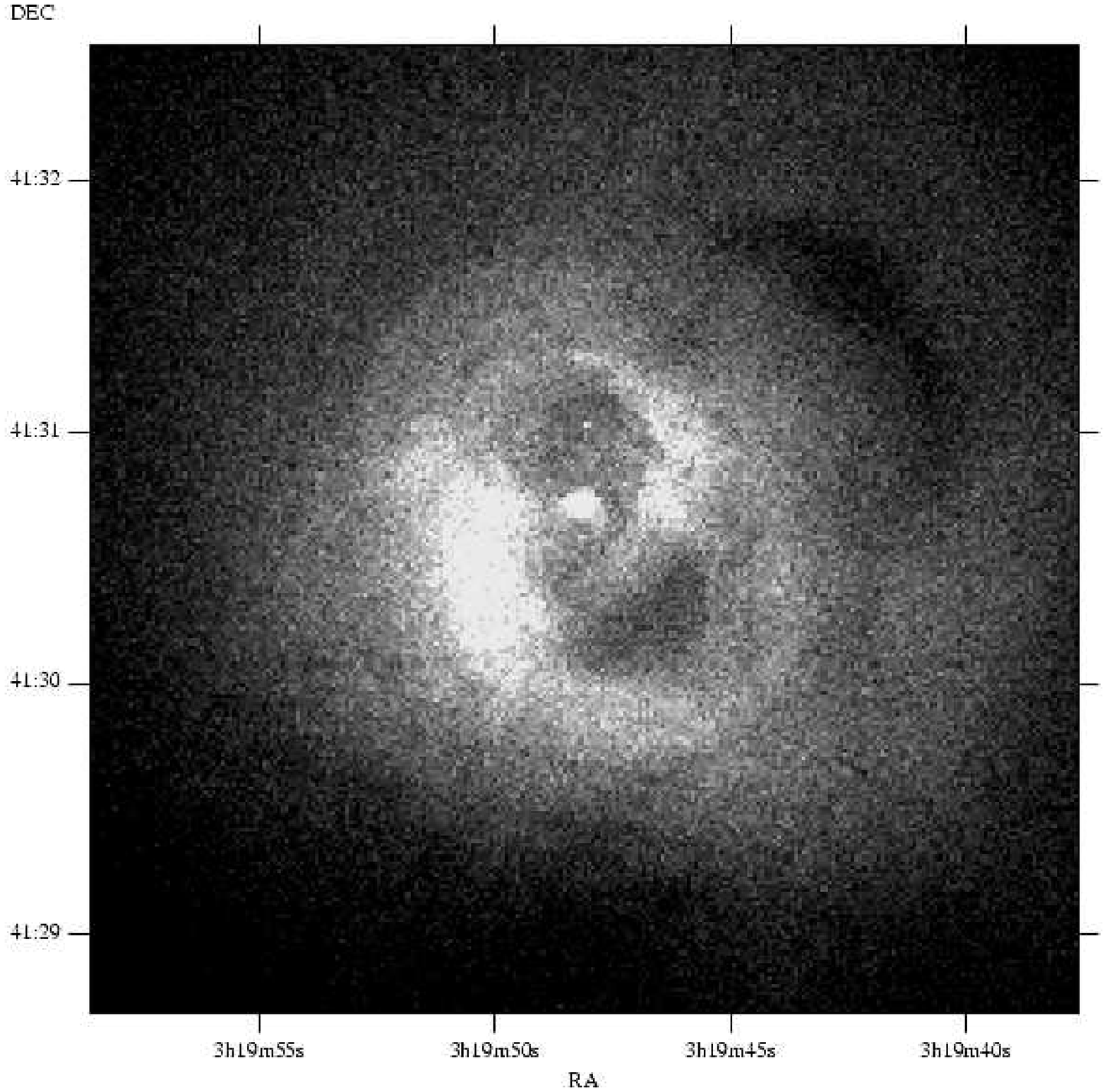}\\
  \includegraphics[width=0.99\columnwidth]{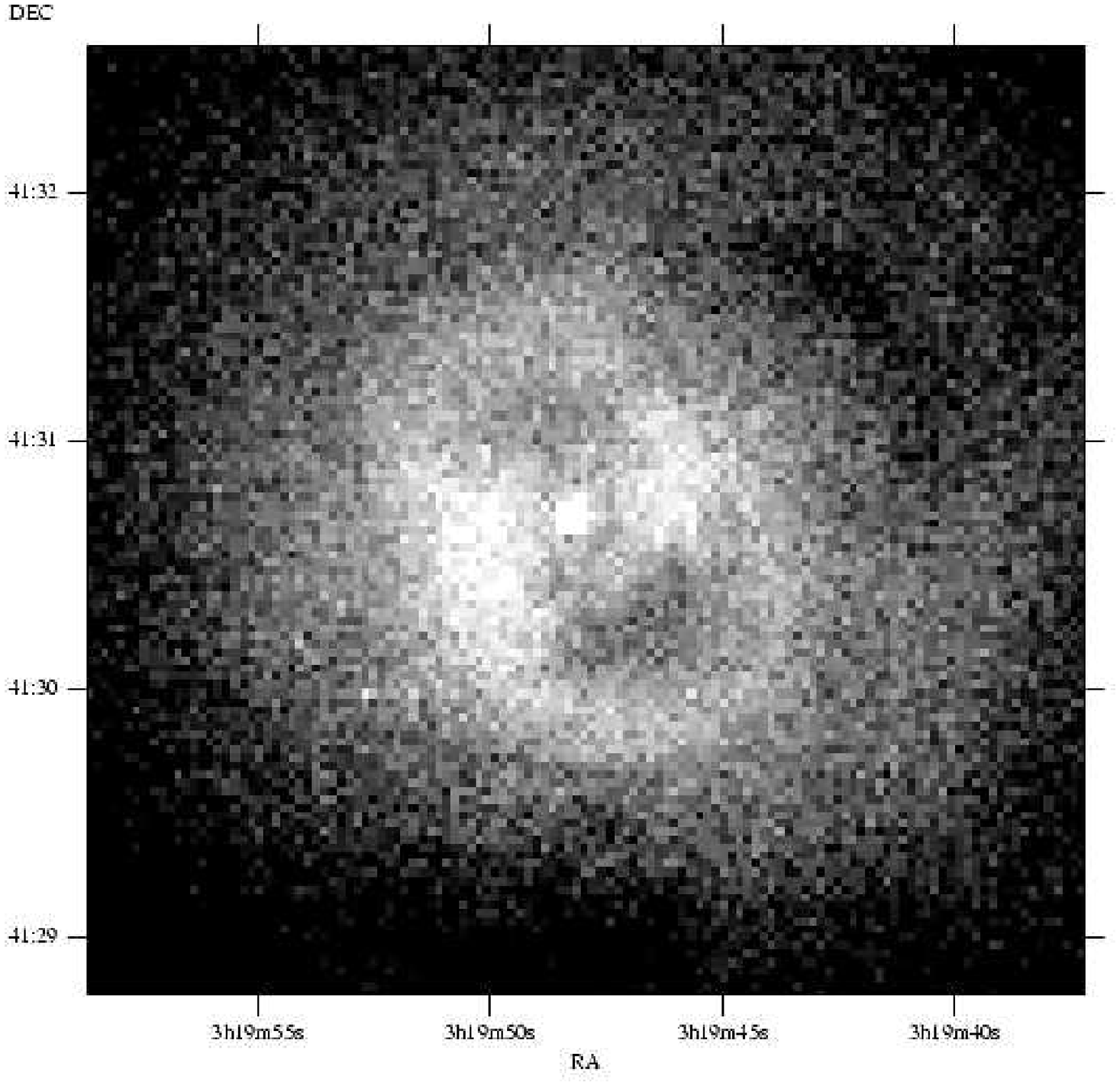}
  \caption{Images in the 1.5--3.5 (upper) and 3.5--7~keV (lower)
bands, binned on 2 and 4 pixels, respectively.}
  \label{fig:asmooth}
\end{figure}

\begin{figure}
  \centering
  \includegraphics[width=0.99\columnwidth]{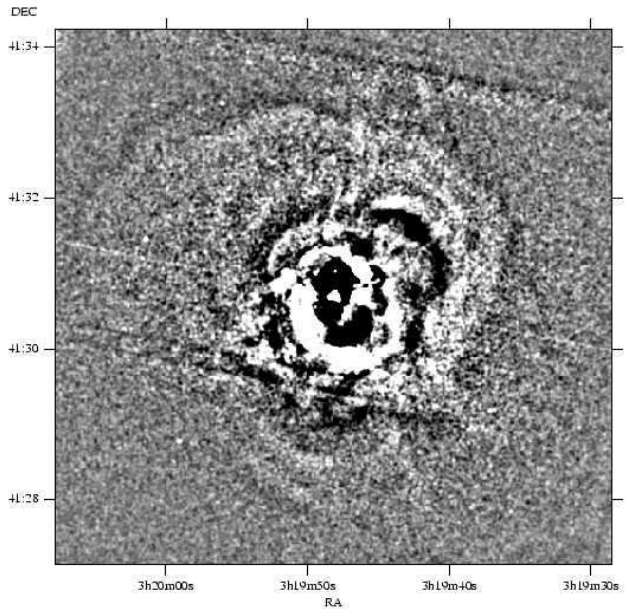}
  \caption{Unsharp-masked image created by smoothing a 0.3 to 7~keV
    intensity image by a $\sigma=9.8$ arcsec Gaussian, subtracting the
    original image, then smoothing by a $0.98$ arcsec Gaussian. The
nodal lines in the detector, mentioned in the text, can be seen
running across the image.}
  \label{fig:unsharp}
\end{figure}

To reveal the larger scale structure in the cluster we used unsharp
 masking. A 0.3-7 keV exposure-map corrected image was smoothed with
 Gaussians of fixed width 0.98~arcsec and 9.8~arcsec, the two smoothed
 images were then subtracted, the result of which is shown in
 Fig.~\ref{fig:unsharp}. Areas in which there were a deficit of counts
 relative to the larger-scale are shown as black, and areas which
 showed a surplus are shaded white. It can be seen that there are a
 number of ripple-like structures which lie outside of the outer radio
 lobes. We have investigated a range of smoothing lengths; the smaller
 one must be less than a few arcsec and the larger one about half a
 ripple wavelength. When their location is known, the ripples can be
 discerned by eye in the original raw image. The implication of the
 ripples is discussed in a later section.

\begin{figure}
  \centering
  \includegraphics[angle=-90,width=0.99\columnwidth]{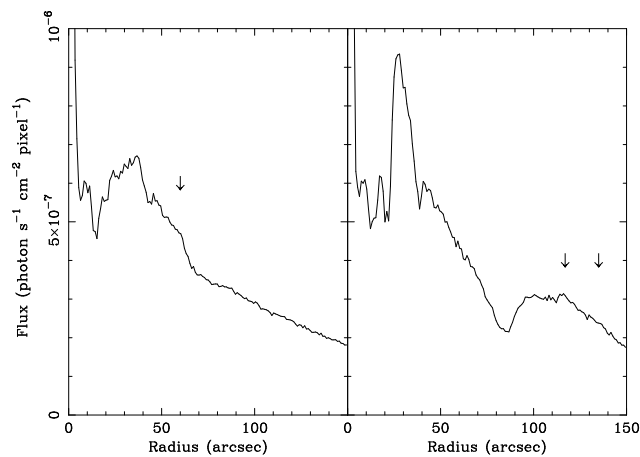}
  \caption{Radial profile in 0.3--7~keV band between p.a. (left)
14--60 deg and (right) 300--320 deg.
Arrows indicate the front (left panel) and ripples
(right panel). Flux uncertainties due to Poisson noise are 1--2 per cent.}
\label{fig:nw_profile}
\end{figure}

To quantify the level of these fluctuations in the unsmoothed image we
generated a radial profile between PAs 300 to $320^{\circ}$, across
the North-West (NW) outer radio lobe. The flux per pixel in
0.98~arcsec wide annuli is shown in Fig.~\ref{fig:nw_profile}. Between
these angles, the positive maxima (relative to the local mean) of the
fluctuations are at radii of $\sim 115$ and $\sim 135$~arcsec. We also
show the profile to the NE, across the interesting, sharp-edged,
emission region just outside rim of the inner N bubble. Its edge, or
front, is at a radius of about 60 arcsec.

\begin{figure}
  \centering \includegraphics[width=0.99\columnwidth]{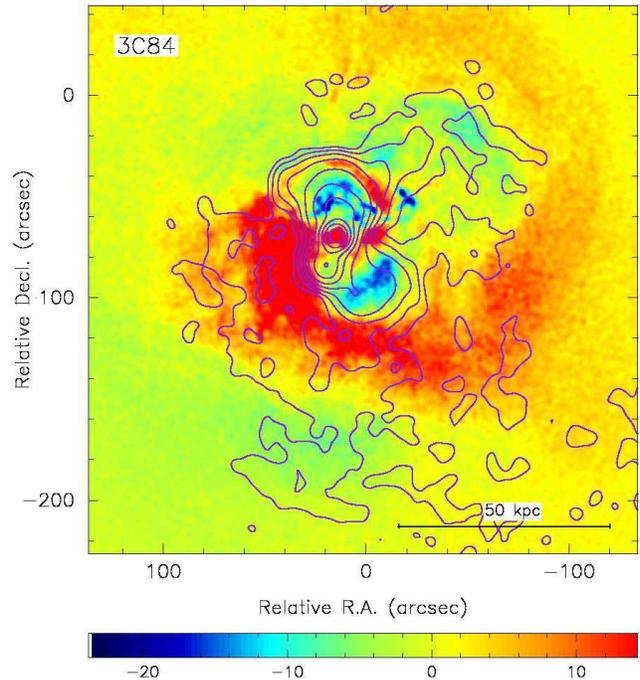}
\caption{Deprojected 0.3 to 7 keV X-ray image, convolved by a 0.98
arcsec Gaussian, overlayed with contours from the 328 MHz VLA image at
a resolution of 8 arcsec.  Contours are drawn at 0.011, 0.022, ...,
and 5.63 Jy/beam.  The peak in the radio image is 8.78 Jy/beam and the
noise is 0.0016 Jy/beam. In addition to filling the inner X-ray
cavities, one can also see a spur of radio emission extending into the
X-ray hole to the NW.} \label{fig:radio}
\end{figure}

In Fig. 5 we show contours from the 328 MHz VLA image.  This image has
been made from combining VLA observations taken in the A, B and C
configurations to produce an image with 8 arcsec resolution and good
sensitivity to extended emission. To produce the deprojected X-ray
image, the average contribution to the emission from outer shells was
subtracted from the pixels within a pixel-width shell, assuming
spherical symmetry around the central source. This procedure
highlights non-radial features in the intensity map.

\begin{figure}
  \centering
  \includegraphics[width=0.99\columnwidth]{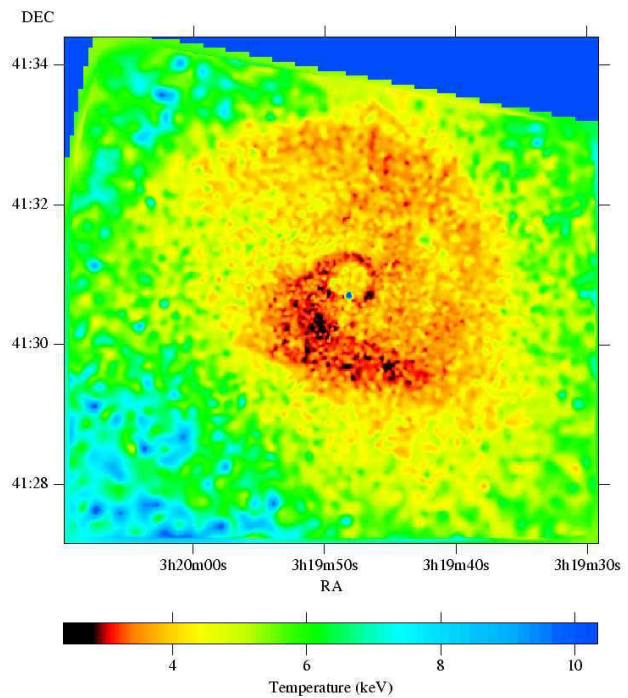}
  \caption{Smoothed temperature map, from spectra in bins with
  2000~counts.} \label{fig:1t_temp}
\end{figure}

To create a temperature map of the cluster, we extracted spectra from
regions containing $\sim 2000$ counts between 0.5 and 7~keV ($S/N \sim
45$), produced using the bin-accretion algorithm of Cappellari \& Copin
(2003). The spectra were binned to contain at least 20 counts per
spectral bin.  The $\sim 6500$ spectra were then fit between 0.5 and
7~keV with a \textsc{mekal} spectral model (Mewe, Gronenschild \& van
den Oord 1985; Liedahl, Osterheld \& Goldstein 1995) absorbed by a
\textsc{phabs} model (Balucinska-Church \& McCammon 1992).  When
fitting the spectra, the temperature, abundance (assuming ratios of
Anders \& Grevesse 1989) and column density were allowed to be free.
Ancillary-response matrices and response matrices were made using the
\textsc{ciao} \textsc{mkwarf} and \textsc{mkrmf} programs, weighting
the responses to regions based on the number of counts between 0.4 and
7~keV.  The ancillary-response matrix was then corrected using the
\textsc{corrarf} routine to correct for the ACIS low energy
degradation\footnote{http://cxc.harvard.edu/cal/Links/Acis/acis/Cal\_prods/qeDeg/},
applying the \textsc{acisabs} absorption profile of Chartas \& Getman
(2002). When fitting, we used a background spectrum generated from a
blank-sky observation in \textsc{caldb} for this period of
observation.


Finally, we made a temperature map (Fig. \ref{fig:1t_temp}) by
smoothing the best-fitting temperatures (accurate to around 10
per~cent, providing the spectral model is correct) using the
\textsc{natgrid} natural-neighbour interpolation
library\footnote{http://ngwww.ucar.edu/ngdoc/ng/ngmath/natgrid/nnhome.html},
assuming the best-fitting temperature to lie at the centroid of each
bin. The temperatures are emission-weighted, and not corrected for the
effects of projection.

In order to demonstrate how the temperature of the cluster changes
over its core, we show in Fig.~\ref{fig:intens_colcode} an intensity
image of the cluster between 0.3 and 7 keV colour-coded using a
temperature map. The temperature map was generated by fitting spectra
to bins containing $10^4$ counts between 0.5 and 7 keV using the same
procedure as above (temperatures accurate from 2 to 4 per~cent). The
lower temperature regions are coded red, rising through green, to the
hotter blue regions (which are hard to see as the brightness of the
object drops dramatically to the outer parts of the image).


\begin{figure}
  \centering
\includegraphics[angle=0,width=0.7\columnwidth,angle=-90]{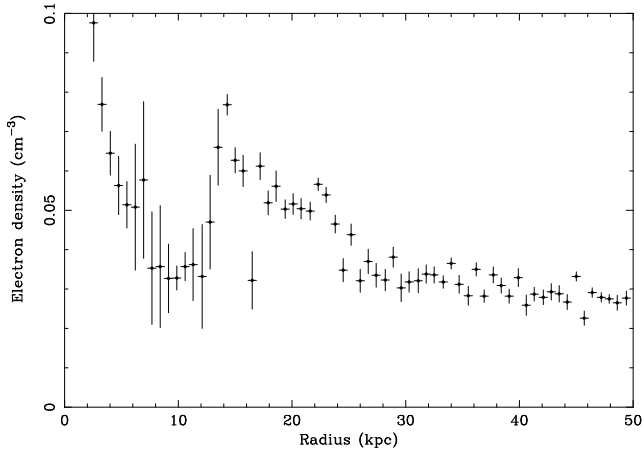}
\caption{Deprojected density profile in a sector to the NW
(pa=14--60 deg).  }
\label{fig:regiont}
\end{figure}

\section{Discussion}
The swirl seen in the gross temperature structure is probably due to
some angular momentum remaining from the merger history of the Perseus
cluster (Fabian et al 2000; Churazov et al 2003; Motl et al 2003).
Otherwise, apart from the ripples discussed below, the gas
distribution beyond the obvious bubbles (2 inner and 2 outer) and
their rims appears to be very smooth. No further bubbles are seen. The
bubbles, discussed by B\"ohringer et al (1993), Churazov et al (2000)
and Fabian et al (2000, 2002b), are inflated by the N-S radio jets from
the nucleus of NGC\,1275. When large enough they detach and rise
buoyantly in the intracluster medium.

The ripples seen in the unsharp-masked image (Fig.~3) are intriguing.
They appear to be quasi-spherical. A prominent inner one to the NE,
immediately outside the northern bubble, has a sharpish edge. The
surface brightness drops by about 20--30 per cent at its edge
(Fig.~4). No obvious temperature or abundance changes occur at this
point, so it is best explained as a density and pressure change. The
amplitude of the first ripple outside the NW hole, which is about 50
per cent further out than the NE one, is much smaller at about 10 per
cent (Fig.~4), and quantifying the other ripples is difficult.

If the ripples are pressure (sound) waves moving at constant speed
(about $1170\kmps$ for a temperature of 5~keV) then their separation
(wavelength) of about 11~kpc means a period of about $9.6\times
10^6\yr$. If the power is constant, the energy flux $\propto
\delta P^2 /n, $ where $P$ and $n$ are the gas pressure and density
respectively, should drop with increasing radius $r$ as $r^{-2}$.
Since the density of the intracluster medium drops approximately as
$r^{-1}$ in this region (Fabian et al 2000) then $\delta P/P$ should
vary as $r^{-1/2}$ and the relative amplitude of ripples as $r^{-1}$
(since the emissivity mainly depends on the density squared). The
observed ripples appear to drop off faster than this, but a much deeper
observation is required to confirm it.

The ripples or sound waves may disperse energy throughout the cooling
region, which is one important criterion for a successful heat source
with which to balance radiative cooling (Johnstone et al 2002; Fabian
2002). An important issue is whether they dissipate energy throughout
this region or not. We have previously assumed that they do not
(Fabian et al 2000). Begelman (2001) and Ruszkowski \& Begelman (2002)
assume that sound waves and small rising bubbles dissipate their
energy on a pressure scale length in their effervescent model for
heating but give no details of the mechanism. Churazov et al (2002)
argue that rising bubbles are responsible for most of the heating.

First we note that ion viscosity can lead to dissipation of the sound
energy within the core. Taking the simple formula of Lamb (1879,
Hydrodynamics, also Landau \& Lifshitz 1959), the distance $L$ over
which the energy of a plane sound wave is reduced by $1/e$ is given by
$${L\over\lambda}={3\over16\pi^2}{{c\lambda}\over\nu}\sim7\lambda_{10}
n_{0.03}T_{5}^{-2},$$ where the wavelength
$\lambda=10\lambda_{10}\kpc,$ sound speed $c$ and viscosity is taken
to be $\nu\sim10^{8} T^{5/2}n^{-1}\cmsq\ps$ (Spitzer 1962, Braginskii
(1958). The ion density $n=0.03n_{0.03}\pcmcu$ and temperature
$T=5\keV$. Thus if the apparent wavelength of about 10~kpc is
maintained then the waves dissipate much of their energy within the
inner 100~kpc of the cooling region. The viscosity assumed here is
high and may of course be significantly affected by magnetic fields in
the gas, so the above estimate is uncertain. The H$\alpha$ filaments
around NGC\,1275 do however provide evidence that the effective
Reynolds number is low and therefore that the viscosity is high
(Fabian et al 2003).

Heating will also result from the initial high amplitude of the sound
waves, which makes them develop rapidly into weak shocks. The sound
speed of the compressed gas is higher (by ${1\over3}{{\delta P}\over
P}$) than that in the rarified part, so the wave will steepen in a few
wavelengths. The sharp edge or front to the NE ripple (Fig.~1, 4 and
8) indicates that this disturbance is already a weak shock.

In order to study the NE front in detail we have looked carefully at
the density profile in this direction. The density jump across the
edge is best determined from a straight surface-brightness
deprojection (e.g. Fabian et al 1981) and is shown in Fig.~7. The
ratio of densities at the edge (inner/outer) is
$1.43^{+0.20}_{-0.13}$. The projected temperature map (Fig.~6) shows
no marked temperature change in this region, although there are a few
wisps of cooler gas (red in Fig.~6). There is therefore a large
pressure increase going inward over the front and it is not a simple
'cold front' (Markevich et al 2000); the inner region is at a higher
pressure and the front must be expanding. This is the condition for a
weak shock to form. Given the above density jump, continuity over the shock
front implies that the temperature should jump by 28 per cent to about
$5.8^{+0.8}_{-0.4}\keV.$ 

To see whether the temperature does jump by this amount, we have
deprojected the spectra from 3 radial zones within a sector across the
front. Assuming a single temperature component within each zone we
find that the temperature drops from $4.19\pm0.06$ to
$3.64\pm0.09\keV$ going inward across the edge. This is inconsistent
with a weak shock, which should heat the gas, unless the gas in the
inner region was already cooler or turbulence behind the shock is
mixing in much cooler gas.

The addition of a some cooler gas in the innermost zone allows for
hotter gas to be present. When added, there is a small reduction in
$\chi^2$, but it is too small to mean that it is statistically
required. We obtain a limit on the volume filling fraction of a
5.8~keV component at 79 per cent (90 percent confidence level).
Despite the high quality of our data we are therefore unable to
measure the temperature jump at the NE front if there is more than a
single (deprojected) temperature component present, even if it
occupies more than half the volume.

Unfortunately this means that we have no temperature confirmation of
the shock. However the sharp increase in density combined with the
temperature measurements does mean that the inner region is at a
higher pressure and should expand outward. The Mach number is 1.5 and
the pressure increases by 84 per cent across the shock.

We note that the edge at the shock front is not perfectly sharp; the
surface brightness decreases over about 5 arcsec or nearly 2~kpc (we
estimate that up to half of this could be due to the point-spread
function of the telescope at this off-axis location). This may
indicate that the mean-free path, $\ell$, is of that order meaning
that the gas is viscous (if $\nu={1\over3}\ell c$). The front is also
seen to the SW in Fig.~1 and shocked gas probably accounts for all the
thick surround to the inner bubbles seen in the hardest image
(Fig.~2b). Projection effects and the multicomponent nature of the
emission preclude making any useful spectral analysis of the hotter
component.

The above discussion assumes purely hydrodynamical processes. Magnetic
fields and cosmic rays in the intracluster gas will complicate many
issues, introducing several more wave modes and also tapping the shock
energy (which could also contribute to the lack of an observed
temperature jump). Particle acceleration will also contribute to the
mini-halo radio source (Gitti et al 2002).  

The rate of work is similar to that required to balance cooling if the
bubbles are formed continuously over a Gyr or more at a mean rate of
one per $10^7\yr$, i.e. the rate deduced from the ripple separation.
If we assume that most of the $PdV$ work done on the surrounding gas
in inflating each bubble propagates as a sound wave, then the $PdV$
energy of about 270 bubbles, each of 7~kpc radius, is needed to
replace the energy radiated by the inner 50~kpc of the cluster, at
which radius the radiative cooling time is $\sim2\times 10^9\yr$. (A
factor of 2 drop in pressure from 12 to 50~kpc is included in this
estimate.) It takes $2.5\times 10^9\yr$ to produce this number of
bubbles if the production rate is on average the same as now.
Therefore the rate of bubble production determined from the
separation/wavelength of the ripples, is close to that required within
50~kpc in order to balance heating and cooling, provided that most of
the sound energy is dissipated within that radius by strong viscosity
and weak shocks. The bubbles, with the relativistic particles and
magnetic fields contained within them, are assumed to rapidly rise
beyond this region and become undetectable.

An important problem raised by the heating mechanism proposed here is
the significant rims of cool gas around the bubbles. They may however
just be a collection of cooler blobs which have been swept up and not
shocked (Fabian et al 2001). Such blobs may be magnetically uncoupled
from the hotter phase, and therefore not conductively evaporated, and
are on a scale which is much smaller than the wavelength of the sound
waves. This means that any pressure differences caused by these waves
across vary more slowly than the internal sound crossing time and no
shock occurs.

In summary, the deep Chandra observation of the Perseus cluster has
revealed subtle ripples which could be the sound waves which transport
and dissipate the energy of the bubbles which continuously form at the
centre. The process is not due to effervescence of many small bubbles
(Begelman 2002), to rare major eruptions (Kaiser \& Binney 2002; Soker
et al 2001) or to the buoyant bubbles themselves (Churazov et al
2001), but to a more continuous dissipation of acoustic energy as the
(attached) bubbles grow. More study is needed to determine whether
further heating is required at larger radii and whether this process
can account for the heating in more luminous objects which (currently)
do not have any major central radio source (e.g. A1835, Fabian et al
2002c; sound waves, generated in the outer hot gas by subcluster
mergers, are refracted inward, Pringle 1989, by the denser core gas,
and may provide the explanation here). The much-sought detailed
connection between the active nucleus, which is possibly fuelled by
the hot surrounding gas, and the required distributed heat source in
the inner intracluster medium has come to light in the Perseus
cluster.

\section*{Acknowledgements}
We thank the referee and Jeremy Goodman for helpful comments and the
Chandra Observatory team for such superb data. ACF, SWA and CSC thank
the Royal Society for support.

\begin{figure*}
  \centering
  \includegraphics[width=2\columnwidth]{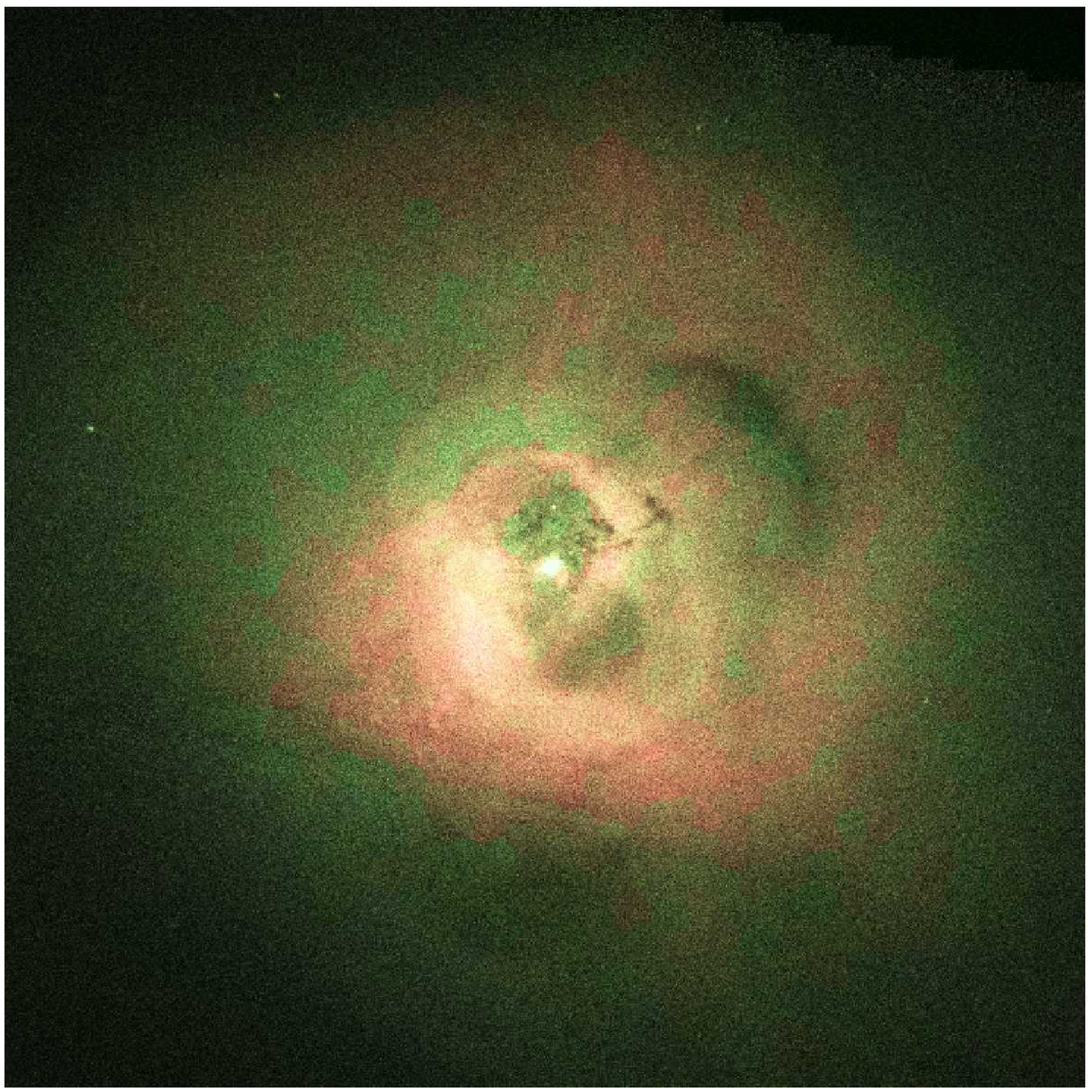}
  \caption{X-ray intensity between 0.3 and 7~keV colour coded by
    temperature (see Section 2). The image is 350 arcsec (131~kpc)
square.}
  \label{fig:intens_colcode}
\end{figure*}


\begin{thebibliography}{}

\bibitem{anders89} Anders E., Grevesse N., 1989, Geochimica et
  Cosmochimica Acta, 53, 197
\bibitem{balchurch92} Balucinska-Church M., McCammon D., 1992, ApJ,
  400, 699
\bibitem  {}  Basson J.F., Alexander P., 2003, MNRAS, 339, 353
\bibitem {} Begelman M.C., 2002, in Gas \& Galaxy
Evolution, ASP Conference Series Vol. 240, J.E. Hibbard, M.P. Rupen
and J.H. van Gorkom, eds. (San Francisco: Astron. Soc. Pacific, 2001),
363
\bibitem  {}  Bertschinger E., Meiksin A., 1986, ApJ, 306, L1
\bibitem  {}  B\"ohringer H., Voges W., Fabian A.C., Edge A.C., Neumann
  D.M., 1993, MNRAS, 264, L25
\bibitem  {}  B\"ohringer H., Matsushita K., Churazov E., Ikebe Y., Chen
  Y., 2002, A\&A, 382, 804
\bibitem  {}  Braginskii S.L., 1958, Sov. Phys., JETP, 6, 358
\bibitem  {}  Brighenti F., Mathews W.G., 2003, ApJ, 587, 580
\bibitem  {}  Br\"uggen M., Kaiser C.R., 2001, MNRAS, 325, 676
\bibitem  {} Cappellari M., Copin Y., MNRAS, 2003, in press, astro-ph/0302262
\bibitem{chartas02} Chartas G., Getman K., 2002,
  http://www.astro.psu.edu/\-users/\-chartas/\-xcontdir/\-xcont.html
\bibitem  {}  Churazov E., Forman W., Jones C., B\"ohringer H., 2000,
  A\&A, 356, 788
\bibitem  {} Churazov E., Br\"uggen M., Kaiser C.R., B\"ohringer H.,
  Forman W., 2001, ApJ, 554, 261
\bibitem{} Churazov E., Sunyaev R., Forman W., B\"ohringer H., 2002,
  MNRAS, 332, 729
\bibitem  {} Churazov E., Forman W., Jones C., B\"ohringer H., 2003,
  submitted to ApJ, astro-ph/0301482
\bibitem  {}  Conselice C.J., Gallagher J.S., Wyse R.F.G., 2001, AJ, 122,
  2281
\bibitem {} Fabian A.C., Hu E.M., Cowie L.L., Grindlay J., 1981, ApJ,
  248, 47
\bibitem  {} Fabian A.C., 1994, ARA\&A, 32, 277
\bibitem  {}  Fabian A.C. et al., 2000, MNRAS, 318, L65
\bibitem  {} Fabian A.C., Fabian A.C., Mushotzky R.F., Nulsen P.E.J.,
  Peterson J.R., 2001, MNRAS, 321, L20
\bibitem  {}  Fabian A.C., Voigt L.M., Morris R.G., 2002a, MNRAS, 335, L71
\bibitem  {}  Fabian A.C., Celotti A., Blundell K.M., Kassim N.E., Perley
  R.A., 2002b, MNRAS, 331, 369
\bibitem  {}  Fabian A.C., Allen S.W., Crawford C.S., Johnstone R.M.,
  Morris R.G., Sanders J.S., Schmidt R.W., 2002c, MNRAS, 332, L50
\bibitem  {}  Fabian A.C., Sanders J.S., Crawford C.S., 
Conselice C.J., Gallagher J.S., Wyse R.F.G., 2003, MNRAS, submitted
\bibitem  {} Gitti M., Brunetti G., Setti G., A\&A, 2002, 386, 456
\bibitem  {} Kaiser C.R., Binney J., 2002, MNRAS, 338, 837
\bibitem {} Johnstone R.M., Allen S.W., Fabian A.C., Sanders J.S.,
2002, MNRAS, 336, 299
\bibitem  {} Lamb H., 1879, Hydrodynamics, Camb. Univ. Press
\bibitem  {} Landau L.D., Lifshitz E.M., 1959, Fluid Dynamics,
Pergamon Press
\bibitem{liedahl95} Liedahl D.A., Osterheld A.L., Goldstein W.H.,
  1995, ApJ, 438, L115
\bibitem  {} Lynds R., 1970, ApJ, 159, L151
\bibitem {} Markevitch M., et al 2000, ApJ, 541, 542
\bibitem{mewe85} Mewe R., Gronenschild E.H.B.M., van den Oord
  G.H.J., 1985, A\&AS, 62, 197
\bibitem  {} Morris R.G., Fabian A.C., 2003, MNRAS, 338, 824
\bibitem  {} Motl P.M., Burns J.O., Loken C., Norman M.L., Bryan G.,
  2003, ApJ, in press, astro-ph/0302427
\bibitem  {} Narayan R., Medvedev M.V., 2001, ApJ, 562, L129
\bibitem  {} Pedlar A., Ghataure H.S., Davies R.D., Harrison B.A.,
  Perley R., Crane P C., Unger S.W., 1990, MNRAS, 246, 477
\bibitem  {} Peterson J. et al., 2001, A\&A, 365, L104
\bibitem  {} Peterson J. et al., 2002, ApJ, astro-ph/0210662 
\bibitem  {} Pringle J.E., 1989, MNRAS, 239, 474
\bibitem  {}  Quilis V., Bower R.G., Balogh M.L., 2001, MNRAS, 328, 1091
\bibitem  {}  Reynolds C.S., Heinz S., Begelman M.C., 2001, ApJ, 549,
  L179
\bibitem{} Ruszkowski M., Begelman M.C., 2002, ApJ, 581, 223
\bibitem  {} Schmidt R.W., Fabian A.C., Sanders J.S., 2002, MNRAS, 337,
  71
\bibitem {} Soker N., White R.E., David L.P., McNamara B.R., 2001, ApJ,
549, 832
\bibitem {} Spitzer L., 1962, The Physics of Fully Ionized Gases,
  Interscience Publishers
\bibitem  {}  Tabor G., Binney J., 1993, MNRAS, 263, 323
\bibitem  {} Tamura T. et al., 2001, A\&A, 365, L87
\bibitem  {} Tucker W., Rosner R., 1983, ApJ, 267, 547
\bibitem  {}  Voigt L.M., Schmidt R.W., Fabian A.C., Allen S.W.,
  Johnstone R.M., 2002, MNRAS, 335, L7
\bibitem  {} 
 Zakamska N.L., Narayan R., 2003, ApJ, 582, 162
\end{thebibliography}
\end{document}